\documentclass[twocolumn,dvipsnames, usenames]{aastex631}
\usepackage[utf8]{inputenc}
\usepackage{amssymb,amsmath,verbatim,mathtools,needspace,enumitem,etoolbox,graphicx,physics,microtype,ulem, breqn}
\usepackage{xspace}
\usepackage{booktabs}
\usepackage{upgreek}
\usepackage{acronym}

\newcommand{\MKI}{\affiliation{MIT Kavli Institute for Astrophysics and Space Research, 70 Vassar St., Cambridge, MA 02139, USA}}
\newcommand{\CfA}{\affiliation{Center for Astrophysics $|$ Harvard \& Smithsonian, 60 Garden Street, Cambridge, MA 02138, USA}}
\newcommand{\Berkeley}{\affiliation{Department of Astronomy, University of California, Berkeley, CA 94720-3411, USA}}
\newcommand{\Caltech}{\affiliation{Division of Physics, Mathematics and Astronomy, California Institute of Technology, Pasadena, CA 91125, USA}}

\acrodef{sn}[SN]{supernova}
\acrodefplural{sn}[SNe]{supernovae}
\acrodef{snia}[SN\,Ia]{Type Ia supernova}
\acrodefplural{snia}[SNe\,Ia]{Type Ia supernovae}
\acrodef{wd}[WD]{white dwarf}
\acrodef{snr}[S/N]{signal-to-noise ratio}
\acrodef{grb}[GRB]{gamma ray burst}

\date{\today}

\begin{document}

\title{A diverse, overlooked population of Type Ia supernovae exhibiting mid-infrared signatures of delayed circumstellar interaction} 

\correspondingauthor{Geoffrey Mo}
 \email{gmo@mit.edu}
\author[0000-0001-6331-112X]{Geoffrey Mo} \MKI
\author[0000-0002-8989-0542]{Kishalay De} \altaffiliation{MIT Kavli Institute Fellow} \MKI 
\author[0009-0002-4843-2913]{Eli Wiston} \Berkeley
\author[0000-0002-8070-5400]{Nayana A.J.} \Berkeley
\author[0000-0003-4768-7586]{Raffaella Margutti} \Berkeley
\affiliation{Department of Physics, University of California, 366 Physics North MC 7300, Berkeley, CA 94720, USA}
\author[0000-0002-7197-9004]{Danielle Frostig} \CfA
\author[0000-0003-1546-6615]{Jesper Sollerman}
\affiliation{The Oskar Klein Centre, Department of Astronomy, Stockholm University, AlbaNova, SE-10691 Stockholm, Sweden}
\author[0000-0003-4531-1745]{Yashvi Sharma}  \Caltech
\author[0000-0003-1169-1954]{Takashi J. Moriya}
\affiliation{National Astronomical Observatory of Japan, National Institutes of Natural Sciences, 2-21-1 Osawa, Mitaka, Tokyo 181-8588, Japan}
\affiliation{Graduate Institute for Advanced Studies, SOKENDAI, 2-21-1 Osawa, Mitaka, Tokyo 181-8588, Japan}
\affiliation{School of Physics and Astronomy, Monash University, Clayton, VIC 3800, Australia}
\author[0000-0002-7226-836X]{Kevin B. Burdge} \MKI
\author[0000-0001-5754-4007]{Jacob Jencson}
\affiliation{IPAC, Mail Code 100-22, Caltech, 1200 E.\ California Blvd., Pasadena, CA 91125, USA}
\author[0000-0003-2758-159X]{Viraj R. Karambelkar} \Caltech
\author[0000-0002-4585-9981]{Nathan P. Lourie} \MKI

\begin{abstract} 
Type Ia supernovae arise from the thermonuclear explosions of white dwarfs in multiple star systems. A rare sub-class of SNe\,Ia exhibit signatures of interaction with circumstellar material (CSM), allowing for direct constraints on companion material. While most known events show evidence for dense nearby CSM identified via peak-light spectroscopy (as SNe\,Ia-CSM), targeted late-time searches have revealed a handful of cases exhibiting delayed CSM interaction with detached shells. Here, we present the first all-sky search for late CSM interaction in SNe\,Ia using a new image-subtraction pipeline for mid-infrared data from the {\it NEOWISE} space telescope. Analyzing a sample of $\approx8500$ SNe\,Ia, we report evidence for late-time mid-infrared brightening in five previously overlooked events spanning sub-types SNe\,Iax, SNe\,Ia-91T and super-Chandra SNe\,Ia. Our systematic search doubles the known sample, and suggests that $\gtrsim 0.05$\% of SNe\,Ia exhibit mid-infrared signatures of delayed CSM interaction. The mid-infrared light curves ubiquitously indicate the presence of multiple (or extended) detached CSM shells located at $\gtrsim 10^{16}-10^{17}$\,cm, containing $10^{-6}-10^{-4}$~$M_\odot$ of dust, with some sources showing evidence for new dust formation, possibly within the cold, dense shell of the ejecta. We do not detect interaction signatures in spectroscopic and radio follow-up; however, the limits are largely consistent with previously confirmed events given the sensitivity and observation phase. Our results highlight that CSM interaction is more prevalent than previously estimated from optical and ultraviolet searches, and that mid-infrared synoptic surveys provide a unique window into this phenomenon.
\end{abstract}


\section{Introduction}
\label{sec:intro}
It is well established that Type Ia supernovae (SNe\,Ia) arise from the thermonuclear explosions of white dwarfs (WDs), triggered by interaction with a companion star \citep{Maoz2014}. Not only do SNe\,Ia serve as standardizable candles \citep{Branch1992, Phillips1993, Kasen2007, Dhawan2018, Avelino2019}, they probe a common end-point in binary stellar evolution \citep{Wang2012, Postnov2014, Ruiter2020, Liu2023}, reveal the fates of mHz gravitational wave sources \citep{Toonen2012, RM2019, Karnesis2021, Korol2024}, and provide insights into the universal chemical and dust budget \citep{Kobayashi2009, Nozawa2011, Lach2020, Wang2024}. However, the nature of the WD (mass; composition) and the companion (degenerate or not; accretion via mergers or mass transfer), as well as the explosion mechanism (detonation or deflagration) remain unresolved, with clear evidence for multiple channels \citep{Pakmor2013, Ashall2016, Polin2019, Bulla2020, Hakobyan2020}. 

Supporting the diverse nature of progenitors, a rare sub-class of SNe\,Ia---known as SNe\,Ia-CSM ($\lesssim 0.2\%$ of SNe\,Ia; \citealt{Silverman2013, Sharma2023})---exhibit strong interaction of the expanding ejecta\footnote{We note that interaction with very nearby CSM can also be detected as early-time excesses in optical light curves or high velocity features (e.g., \citealt{Noebauer2016, Mulligan2017, Moriya2023}); however, we do not discuss them here since existing observations do not conclusively rule out other possibilities (e.g. \citealt{Childress2014, Deckers2022}).} with dense nearby circumstellar material (CSM). These events are typically identified via strong narrow emission lines (commonly H$\alpha$) in spectroscopy acquired at or near peak light \citep{Sharma2023}, similar to Type IIn supernovae \citep{Smith2017_handbook}. This H-rich material has been ubiquitously attributed to mass supplied by a non-degenerate companion, allowing for direct constraints on its composition and configuration. The proximity ($\lesssim 10^{15}$\,cm) and large mass ($\gtrsim 0.1$\,M$_\odot$) of the CSM have been explained via binary progenitors such as a WD merging with the core of an asymptotic giant branch (AGB) star inside (or shortly after the ejection) of the common envelope (i.e. the core-degenerate scenario; \citealt{Tutukov1992, Livio2003, Kashi2011, 2021PASP..133g4201A}) or even as a result of dynamical mass transfer from a main sequence star \citep{Han2006}. Since systematic searches \citep{Silverman2013, Sharma2023} have relied on peak-light spectroscopy or peculiar undulations in early-time light curves to identify candidates for SNe\,Ia-CSM, these are biased against SNe\,Ia with distant CSM shells where signatures of interaction would only be detectable at later phases \citep{Sharma2023, Phillips2024}.

\citet{Dilday2012} presented the first example of a SN\,Ia (of the 1991T subclass; see \citealt{Blondin2012} and \citealt{Taubenberger2017} for a review of SN\,Ia sub-types) that transitioned to a Ia-CSM at late phases ($> 40$\,d after peak) due to the delayed CSM interaction in PTF11kx.\footnote{\citet{Harris2018} names these SNe\,Ia;n to indicate the late-time emergence of narrow (``n'') emission lines.} The multiple, detached CSM shells detected in this object were suggested to arise from a symbiotic nova progenitor \citep{Han2004}, although \citet{Soker2013} have argued for a core-degenerate scenario where the envelope was ejected shortly before the explosion. Using a \textit{Hubble Space Telescope (HST)} ultraviolet photometric survey of 72 SNe, \citet{2019ApJ...871...62G} reported evidence for late CSM interaction in SN\,2015cp, which was initially classified as a SN\,Ia-91T-like object. A search for similar emission in the GALEX archive yielded no additional detection \citep{Dubay2022}. Late-time spectroscopic signatures of interaction (via H$\alpha$ emission) have also been reported in SN\,2016jae \citep{2021A&A...652A.115E}, SN\,2018cqj \citep{2020ApJ...889..100P}, and SN\,2018fhw \citep{2019MNRAS.486.3041K, 2019MNRAS.487.2372V}, while \citet{Kool2023} recently reported the first case of a SN\,Ia interacting with He-rich CSM. While systematic efforts have been previously made to identify potential candidates for late interaction with optical photometry alone \citep{Terwel2024}, developing a complete census either by continued spectroscopic monitoring or UV observations is challenging on large scales due to cost.

The mid-infrared (MIR) bands provide an independent and powerful approach to probing CSM interaction and associated dust formation \citep{Szalai2019}. Instead of directly detecting the high-energy X-ray or UV radiation produced by the shock interaction with the CSM, MIR observations detect emission from warm dust, which can either be pre-existing around the progenitor or be produced in the interaction region after explosion \citep{Fox2011, Szalai2013}. When SN ejecta interact with surrounding CSM, dust can be heated by shock radiation, producing luminous late-time MIR emission \citep{Fox2015}. MIR brightening episodes therefore provide a unique tool to detect CSM interaction at late phases, but systematic searches remain limited without synoptic MIR capabilities. \citet{Myers2024} have recently reported a systematic search for late MIR emission in core-collapse supernovae from \textit{NEOWISE} data \citep{Wright2010, Mainzer2014}, demonstrating the potential of all-sky, slow cadence surveys in revealing the demographics of CSM interaction.

In this Letter, we report the identification of an overlooked population of recent SNe\,Ia exhibiting delayed CSM interaction. Capitalizing on the rapid growth of large SN samples from all-sky time-domain optical surveys such as ZTF \citep{Bellm_2018}, ATLAS \citep{Tonry2018}, ASAS-SN \citep{Kochanek2017}, Pan-STARRS \citep{Chambers2016}, and Gaia \citep{gaia_alerts}, we present a complete search of \textit{NEOWISE} data for MIR excesses using a new image subtraction pipeline. We begin by characterizing the MIR behavior of known SNe\,Ia with delayed CSM interaction, and a search for new similar objects in Section~\ref{sec:data}. We use the MIR emission together with follow-up observations to constrain the amount and configuration of the CSM shells in Section~\ref{sec:dust}. We conclude by discussing the implications for this overlooked population in Section~\ref{sec:discussion}. Throughout, we assume \texttt{Planck18} cosmology \citep{Planck2018}.

\section{Observations}
\label{sec:data}

\subsection{MIR behavior of known SNe\,Ia with late CSM interaction}
\label{subsec:known}

\begin{figure*}[!htbp]
    \centering
    \includegraphics[width=\linewidth]{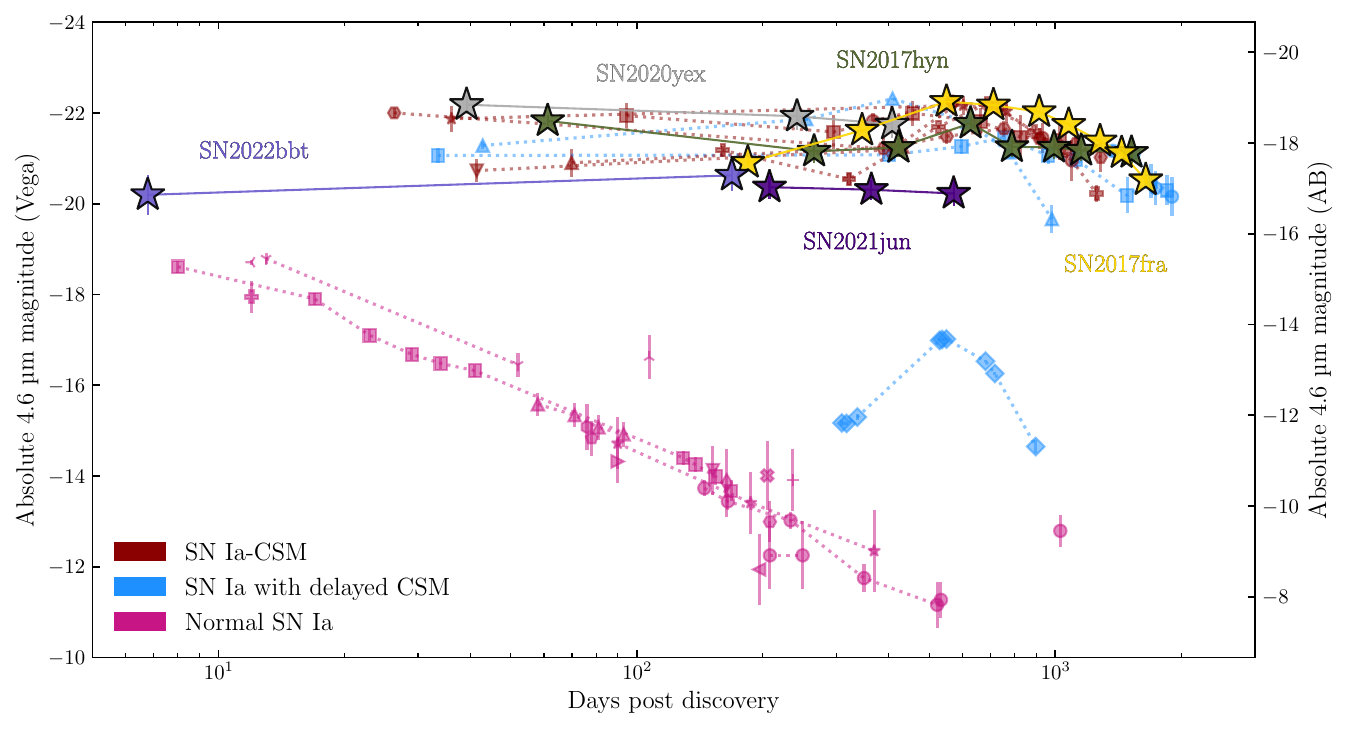}
    \caption{The phase space of mid-infrared light curves for SNe\,Ia.
    We show \textit{NEOWISE} $W2$ light curves for the SNe\,Ia showing delayed CSM interaction from our sample in the colored stars.
    For comparison, we also plot previously published SNe\,Ia with delayed CSM interaction, including PTF11kx, SN\,2015cp, and SN\,2020eyj (blue circles, squares, and triangles respectively).
    Plotted in the blue diamonds is the {\it Spitzer} IRAC Ch2 light curve of SN\,2014dt, a SN\,Iax which exhibits an analogous rise and fall to the interacting SNe\,Ia, but at a much lower luminosity \citep{2016ApJ...816L..13F, Jencson2019}.
    The SNe\,Ia-CSM, for which we show selected $W2$ light curves in dark red (these include SNe\,2016iks, 2017eby, 2017hzw, 2018crl, 2018evt, 2018gkx, 2019agi, 2020aekp), display similar behavior to most of the SNe\,Ia with delayed CSM interaction.
    Finally, we show {\it Spitzer} Ch2 light curves of normal SNe\,Ia from \citet{Szalai2019} in magenta.}
    \label{fig:mir_lcs}
\end{figure*}

Using data from the infrared {\it Spitzer Space Telescope} \citep{Spitzer}, \citet{Fox2013} reported luminous, brightening late-time MIR emission in SNe\,Ia-CSM 2002ic and 2005gj, suggesting renewed shock interaction with CSM shells. Subsequently, \citet{Szalai2019}, \citet{Szalai2021} and \citet{Sharma2023} also confirmed consistently bright late-time MIR emission in larger samples of SNe\,Ia-CSM based on {\it Spitzer} and \textit{NEOWISE} observations, largely consistent with pre-existing dust heated by ongoing CSM interaction. A similar, luminous MIR excess was also reported in the case of the nearby SN\,Iax 2014dt \citep{2016ApJ...816L..13F, Jencson2019}, although its origin has been debated (as arising from circumstellar dust or possibly a bound remnant; \citealt{Foley2016}). Using a comprehensive spectral sequence and \textit{NEOWISE} observations, \citet{Wang2024} provided evidence for multiple dust shells as well as new late-time dust formation in the SN\,Ia-CSM 2018evt. To understand if MIR searches can be fruitful to look for delayed CSM interaction in SNe\,Ia, we first investigate previously known events in \textit{NEOWISE} data. The \textit{NEOWISE} survey's all-sky coverage, MIR photometric bands ($W1 \approx 3.4\,\mu$m, $W2 \approx 4.6\,\mu$m), long baseline ($\approx 13$\,yr), regular cadence ($\approx 6$\,month), and sensitivity ($\approx 20$\,AB mag) are ideal attributes for such a search. When combined with a new difference photometry pipeline (De et al., in prep) with unWISE coadded images \citep{Lang2014, Meisner2018} and a ZOGY-based subtraction algorithm \citep{Zackay2016, De:2019xhw} adept at recovering faint transients on bright hosts, this MIR dataset has already proven to be powerful for core-collapse SNe \citep{Myers2024}.

\textit{NEOWISE} observations of PTF11kx begin $\approx 3.5$\,yrs after peak light, exhibiting luminous emission detected for two more years (also previously reported with \textit{Spitzer} data in \citealt{2017ApJ...843..102G}). Similarly, the \textit{NEOWISE} analysis of SN\,2015cp reveals a long-lasting MIR light curve\footnote{\citet{Thevenot21RNAAS} previously also reported the \textit{NEOWISE} detection of SN\,2015cp.}. We show a comparison of these light curves with SNe\,Ia-CSM in Figure \ref{fig:mir_lcs}, and their individual light curves in Figure~\ref{fig:old_lcs}. \citet{Kool2023} reported luminous MIR emission from SN\,2020eyj that exhibited delayed interaction with a He-rich CSM shell. In contrast, \textit{NEOWISE} light curves of the sub-luminous SNe\,Ia which show late-time H$\alpha$ emission (SNe\,2016jae, 2018cqj, and 2018fhw), as well as the candidates identified from late-time optical photometry \citep{Terwel2024} (SNe\,2018grt, 2019ldf, and 2020tfc) do not show luminous MIR emission comparable to the other events. We discuss possible reasons in Section~\ref{subsec:multiwavelength}. As shown in Figure~\ref{fig:mir_lcs}, many SNe\,Ia-CSM and SNe\,Ia with delayed CSM interaction show luminous late-time MIR emission, distinct from normal SNe\,Ia, making MIR searches a promising avenue for identifying this phenomenon.

\subsection{Systematic search in NEOWISE for missed events}
\label{subsec:search}

\begin{figure*}[!ht]
    \centering
    \includegraphics[width=\linewidth]{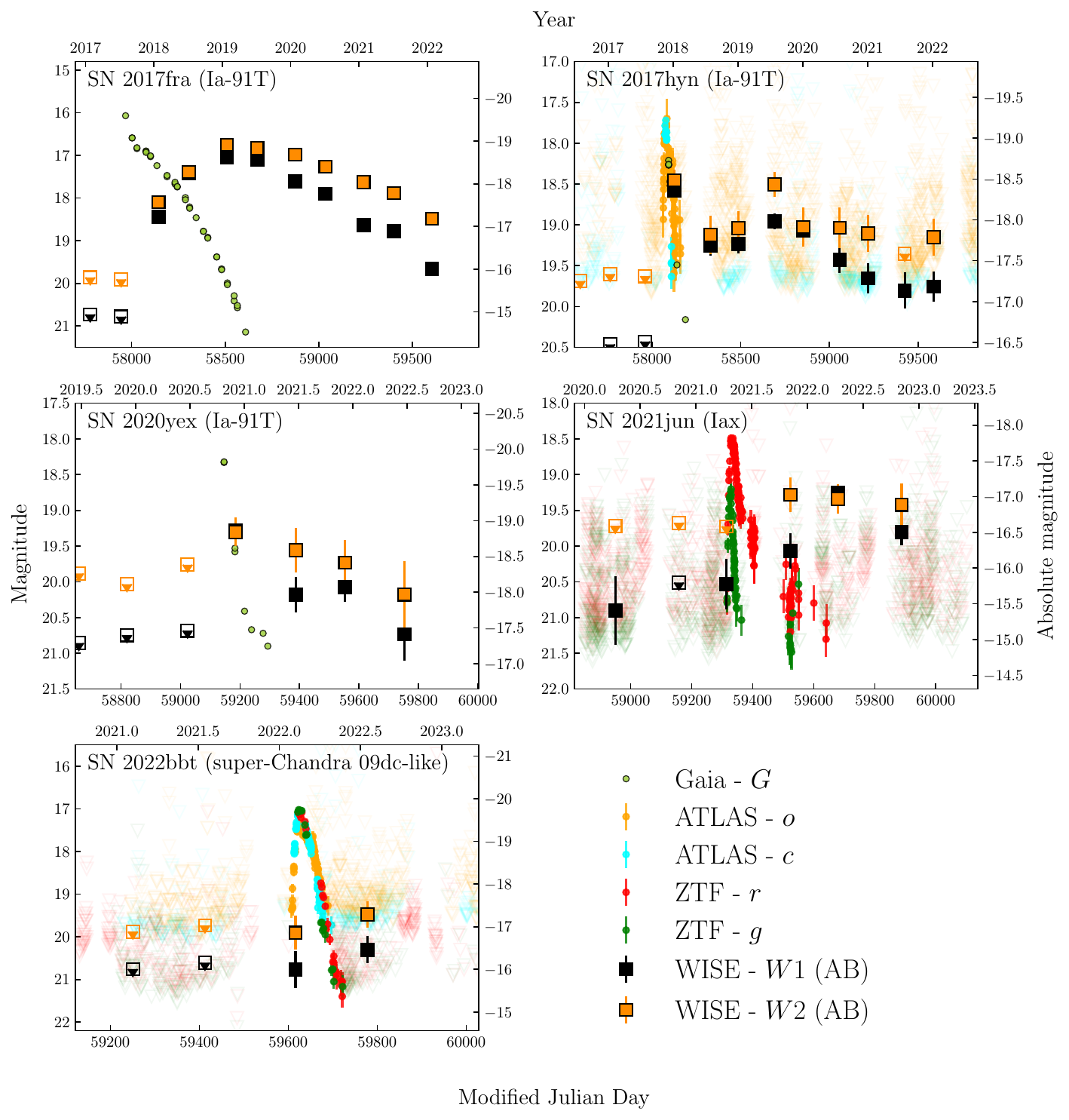}
    \caption{Optical and mid-infrared light curves of our sample of Type Ia SNe showing evidence for delayed CSM interaction.
    We show \textit{NEOWISE} MIR light curves alongside Gaia, ATLAS, and ZTF photometry of the SNe in our sample.
    Upper limits are displayed as downward pointing empty markers.
    In parentheses next to each SN's label is its peak-light classification.
    \label{fig:lcs}}
\end{figure*}

We produced \textit{NEOWISE} MIR difference imaging light curves for $\approx 8500$ Type Ia SNe from 2011 to 2022\footnote{The 2022 cutoff is governed by the availability of unWISE data products.} reported to the IAU Transient Name Server\footnote{\url{https://www.wis-tns.org}}. These light curves were visually inspected to search for late-time MIR emission associated with the SN. We filter the light curves by i) rejecting candidates with detections prior to the optical onset of the SN (to remove contamination from active galactic nuclei activity) or obvious subtraction artifacts; and ii) requiring at least two epochs of \textit{NEOWISE} detections ($\approx 1$\,yr) and rising emission in at least one band, to conclusively rule out excess MIR emission arising purely from new dust formation or a SN light echo without added power from shock-interaction, and to eliminate detections deemed to only be associated with photospheric emission from the optical SN. Most of the identified candidates were known SNe\,Ia-CSM; we do not discuss them here as their light curves have been presented in \citet{Sharma2023}. Table~\ref{tab:cand_cuts} summarizes the cuts at each filtering stage. Notably, these selection criteria resulted in the identification of five candidates which exhibit luminous late-time MIR emission despite not being previously associated with CSM interaction.\footnote{SN\,2020mvp \citep{2020TNSTR1827....1T} also emerged as a candidate, but further astrometric inspection of the \textit{NEOWISE} imaging revealed that the MIR emission was likely from temporally coincident nuclear activity.}

The candidates span multiple Type Ia subclasses: SN\,2017fra (originally classified as a SN\,Ia-91T), SN\,2017hyn (Ia-91T), SN\,2020yex (Ia-91T), SN\,2021jun (Iax), and  SN\,2022bbt (super-Chandra Ia)\footnote{\citet{Thevenot21RNAAS} and \citet{Thevenot21Astronote} have previously noted MIR detections for SN\,2017hyn, SN\,2017fra, and SN\,2020yex; \citet{2021ApJS..252...32J} previously analyzed \textit{NEOWISE} observations of SN\,2017fra in their search for MIR flares in nearby galaxies.}. We summarize these SNe in Table \ref{tab:sne}, further describe each in Appendix~\ref{app:sample}, and show reference, science, and difference image cutouts in Figure~\ref{fig:ref_stack_sub}. While half of our sample are SNe\,Ia-91T-like, strengthening the established connection between SNe\,Ia-CSM and SNe\,Ia-91T \citep{2015A&A...574A..61L}, we report the first detection of late-time MIR brightenings in a super-Chandra SN\,Ia,\footnote{The super-Chandra SNe\,Ia SN\,2012dn \citep{Nagao2017} and SN\,2022pul \citep{Siebert2024} have previously reported MIR excesses, but the lack of a detected MIR brightening are consistent with SN light echoes or new dust formation in cooling ejecta.} and the second detection of an MIR excess in a SN\,Iax. We show a comparison of their MIR light curves in Figure~\ref{fig:mir_lcs}, and the individual optical and MIR light curves of these events in Figure~\ref{fig:lcs}. Compared to normal SNe\,Ia, our newly identified events have significant MIR excesses reaching luminosities of $W2 \approx -22$\,Vega\,mag, lasting hundreds to thousands of days. Of note is the MIR light curve of the SN\,Iax 2014dt, which shows similar behavior to but is much less luminous than the rest of the interacting SNe\,Ia; its late MIR rise has been suggested to arise from newly formed dust, but could also be attributable to pre-existing CSM or a bound remnant \citep{2016ApJ...816L..13F, Foley2016, Jencson2019}.

\begin{table*}[!htpb]
      \caption{Summary of our SNe Ia showing MIR signs of delayed CSM interaction. The integrated energy is calculated by integrating the detected W2 fluxes.}
    \begin{tabular}{lllllll}
    \toprule
      Name & Type & Host morphology & Redshift & Distance & Peak W2 & Integrated W2\\
      & & & & [Mpc] & AB magnitude & energy [erg]\\
          \midrule
          \midrule
    SN\,2017fra & Ia-91T & Interacting spiral & 0.03 & 136 & 16.8 & $7.9 \times 10^{49}$ \\
    SN\,2017hyn & Ia-91T & Edge-on disk & 0.053 & 244 & 18.5 & $5.4 \times 10^{49}$ \\
    SN\,2020yex & Ia-91T & Blue spheroidal & 0.087 & 425 & 19.3 & $2.3 \times 10^{49}$ \\
    SN\,2021jun & Iax    & Blue spheroidal & 0.040 & 183 & 19.3 & $5.2 \times 10^{48}$ \\
    SN\,2022bbt & Ia-SC  & Face-on spiral & 0.049 & 225 & 19.5 & $2.8 \times 10^{49}$ \\
          \bottomrule
    \end{tabular}
    \label{tab:sne}
\end{table*}

\subsection{Multi-wavelength follow-up observations}
\label{subsec:followup}
We performed optical spectroscopic follow-up of two of the youngest events\footnote{We did not follow up SN\,2021jun due to its proximity to its host nucleus, which would make faint object spectroscopy challenging.}: SN\,2020yex ($z = 0.087$) and SN\,2022bbt ($z = 0.049$). We also performed radio follow-up of SN\,2022bbt. Optical spectroscopic observations were carried out using a Fast Turnaround Program on the Gemini North and Gemini South telescopes (Program IDs GN-2024B-FT-102/GS-2024B-FT-102; PI: Mo), while radio observations were carried out using the Karl G. Jansky Very Large Array (VLA) under the Director's Discretionary Program (VLA/24A-498; PI: Nayana AJ). In addition, we obtained NIR spectroscopy of SN\,2020yex and NIR imaging of SN\,2017fra using the Magellan Baade telescope, and NIR imaging of SN\,2021jun with {\it WINTER} \citep{Lourie2020, frostig2024winter}. Further details about the observations and data reduction can be found in Appendix~\ref{app:obs}. We do not detect photometric or spectroscopic signatures of CSM interaction in any of our optical/NIR observations while the VLA observations also did not detect a radio counterpart. We discuss the implications of these non-detections in Section~\ref{subsec:multiwavelength}.

\section{Origin of the MIR emission}
\label{sec:dust}
We have identified a sample of five SNe\,Ia showing clear evidence for late-time MIR brightenings. While fading excess late-time MIR emission can be caused by a light echo from the SN or dust formation in the cooling ejecta, rebrightening in the MIR light curve indicates the presence of an internal energy source (i.e., high-energy radiation in the form of UV or X-ray photons) that is heating pre-existing or newly-formed dust \citep{Fox2010, Fox2011}. The lack of observed UV or X-ray emission from these events is unsurprising given the non-existence of all sky UV/X-ray surveys and the expected luminosity of the shock interaction radiation. For the observed peak W2 MIR luminosity of $\sim 10^{42}$\,erg\,s$^{-1}$, the equivalent peak X-ray flux for sources at $\gtrsim 200$\,Mpc would be $\sim 10^{-13}$\,erg\,cm$^{-2}$\,s$^{-1}$, which is far below the detection threshold of existing X-ray all-sky surveys ($\sim 10^{-11}$\,erg\,cm$^{-2}$\,s$^{-1}$ for MAXI/GSC, \citealt{2013ApJS..207...36H}; $\sim 10^{-10}$\,erg\,cm$^{-2}$\,s$^{-1}$ for Swift/BAT, \citealt{2013ApJS..209...14K}), and may be detectable only with pointed X-ray follow-up (which does not exist for these sources). Furthermore, targeted X-ray observations of other SNe\,Ia have had limited success due to the rarity (one known event in SN\,2012ca; \citealt{2018MNRAS.473..336B}) and low luminosity ($L_X \lesssim 10^{37}$\,erg\,s$^{-1}$ for normal SNe\,Ia from \citealt{2014ApJ...790...52M}; $L_X \lesssim 10^{40}$\,erg\,s$^{-1}$ for SNe\,Ia with delayed CSM interaction from \citealt{2023MNRAS.520.1362D, 2024MNRAS.533...27D}) of X-ray emission associated with SNe\,Ia.

Given that known SNe\,Ia exhibiting delayed CSM interaction (see Figure~\ref{fig:mir_lcs}) also show late-time MIR brightenings, these observations strongly suggest that the brightenings are powered by the onset of CSM interaction, where the shock of the high-velocity SN ejecta impacting existing CSM creates high-energy radiation that is absorbed and reemitted by dust. The integrated MIR energies shown in Table~\ref{tab:sne} span approximately 1-10\% of the $\sim10^{51}$\,erg in kinetic energy released in a SN\,Ia due to the efficient conversion of kinetic energy to radiative energy during the shock; these shocks can convert up to $\sim 50\%$ of the total kinetic energy into radiative energy for interacting core-collapse SNe with similar CSM interaction \citep{Smith2017_handbook}. The lack of spectroscopic signatures or optical light curve abnormalities at peak-light indicate that this sample represents an overlooked population of SNe\,Ia undergoing delayed CSM interaction. In this section, we use the \textit{NEOWISE} multi-band photometry to constrain the mass, temperature, and location of the emitting dust to infer the CSM properties. We assume a spherical configuration for the dust for simplicity, though our limited observations cannot rule out complex geometries which may affect the light curve or inferred parameters only up to order unity corrections. In addition to the five SNe, we also fit the \textit{NEOWISE} data from SN\,2015cp, for comparison to a well-studied event.

\subsection{Dust modeling}
\begin{figure}
    \centering
    \includegraphics[width=\linewidth]{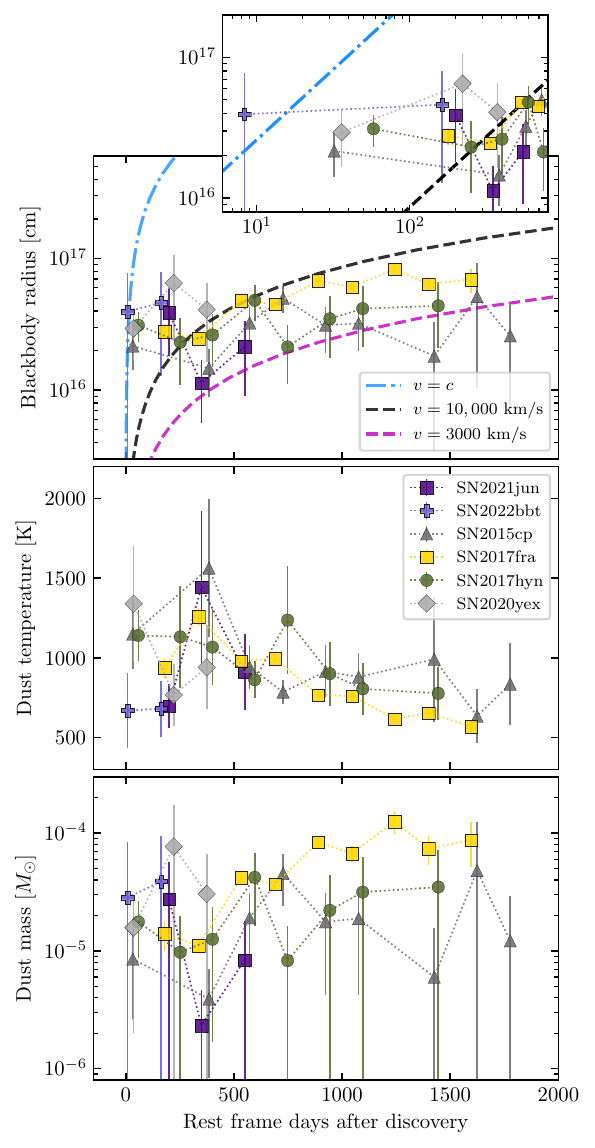}
    \caption{The evolution of dust properties, derived from fitting the \textit{NEOWISE} $W1$ and $W2$ bands for events in our sample, as well as for SN\,2015cp.
    \textit{Top:} Lower limits on the blackbody radii compared to the speed-of-light (blue line), $10^4$ km\,s$^{-1}$ ejecta (black line; relevant for typical SNe\,Ia), and 3000~km\,s$^{-1}$ ejecta (magenta line; relevant for SNe\,Iax).
    Note that the dust radii at early phases are far outside the ejecta, as shown in the inset.
    \textit{Middle:} Dust temperatures assuming 0.1\,$\mu m$ carbonaceous grains. There is a rise in the dust temperature at $\approx 1$\,yr after explosion, indicating the reheating of pre-existing dust by radiation produced by CSM interaction.
    \textit{Bottom:} The evolution of the dust mass shows an increase at later phases, consistent with the formation of new dust.}
    \label{fig:dust}
\end{figure}

We first fit the \textit{NEOWISE} photometry to a pure blackbody, described by
\begin{equation}
    F_\nu = \frac{r_{\rm bb}^2}{D^2} \pi B_\nu(T_{\rm bb}),
    \label{eq:bb}
\end{equation}
where $F_\nu$ is the observed spectral flux density, $r_{\rm bb}$ is the blackbody radius, $D$ is the distance to the source, and $T_{\rm bb}$ is the blackbody temperature. Given that the dust is not expected to emit as a pure blackbody source and may be optically thin, the derived $r_{\rm bb}$ is a lower limit to the true radius of the dust shell. Assuming the dust to be optically thin, the dust mass and temperature are described by
\begin{equation}
    M_d = \frac{4 a \rho D^2 F_\nu}{3 Q_\nu \pi B_\nu(T_d)},
    \label{eq:dust_model}
\end{equation}
where $M_d$ is the dust mass, $a$ is the dust size, $\rho$ is the mass density of the dust, $Q_\nu$ is a dimensionless quantity describing the dust emissivity at frequency $\nu$ for dust size $a$, $B_\nu$ is the Planck function, and $T_d$ is the dust temperature \citep{1977ApJ...216..698H}. We assume a dust grain size of 0.1\,$\mu $m, a composition of carbonaceous grains, a dust mass density of $\rho = 1.0$\,g\,cm$^{-3}$, and we use the emissivity function for graphite dust from \citet{2001ApJ...551..807D}. Fits with different dust composition (silicate vs graphite), grain size (0.01\,$\mu$m vs 0.1\,$\mu$m), and dust mass density (1\,g\,cm$^{-3}$ vs 2.2\,g\,cm$^{-3}$ from \citealt{2007ApJ...657..810D} vs 5\,g\,cm$^{-3}$) resulted in similar results, with dust masses within factors of a few and consistent dust temperatures. Since this fit assumes optically thin dust, and we lack observations at longer wavelengths which may help better constrain the dust, these are lower limits on dust masses. The evolution of our fit parameters is shown in Figure~\ref{fig:dust}.

\subsection{Constraints on CSM configuration}
The top panel of Figure~\ref{fig:dust} shows the blackbody radius evolution compared to the ejecta radius of the SN. For all events, the blackbody radius is outside the ejecta radius at early times ($<300$\,d), indicating that the emission at those times is powered by pre-existing dust. With an early SN luminosity of $\sim 10^{42} - 10^{43}$\,erg\,s$^{-1}$, the vaporization radius for graphite-rich $0.1\,\mu$m grains is a few $\times 10^{16}$\,cm \citep{Fox2010}, consistent with the survival of dust at the derived scales. Figure~\ref{fig:dust} also shows that the radii of the largest dust shells are comparable to the light travel time for the epoch of observation, suggesting that pre-existing dust is likely immediately re-radiating the optical SN light as a dust echo. A dust shell with a radius of a few $\times 10^{16}$\,cm dictates that an observed dust echo cannot last longer than $\sim 100$\,d given the typical duration of a SN\,Ia and the light travel time across the shell. In addition, the available peak-light spectra do not show evidence of CSM interaction, indicating that the early-onset MIR emission is likely powered by the SN light itself.

Throughout our sample, we observe continued MIR emission beyond $\gtrsim 100$\,d, with a subsequent rise in the MIR light curve at $\approx 1-2$\,yrs after explosion that is accompanied by a rise in the dust temperature. This indicates that pre-existing dust is being re-heated by the SN ejecta interacting with existing CSM at $\sim10^{16}$\,cm, assuming a typical ejecta velocity of $10^4$\,km\,s$^{-1}$. Given that the pre-existing dust is typically at a larger radius, the observations suggest the presence of multiple CSM shells, as inferred for both SNe\,Ia-CSM \citep{Fox2013, Wang2024} and previous SNe\,Ia displaying delayed CSM interaction \citep{Dilday2012}. However, we cannot rule out the possibility that the SNe are surrounded by single, detached thick shells (with outer-to-inner radius ratio $\gtrsim $few). At early times, we additionally constrain the pre-existing dust mass to be $\sim 10^{-6} - 10^{-4}$\,M$_\odot$ in all the events.

The subsequent emission at later times ($\gtrsim 2$\,yr), in events for which we have well-sampled late-time MIR light curves, have blackbody radii within the ejecta radius together with a slow increase that traces the ejecta radius evolution (SN\,2015cp, SN\,2017fra, and SN\,2017hyn). When combined with the potentially increased dust mass, this suggests the formation of new dust, possibly within the cold, dense shell of the ejecta as it interacts with existing CSM, consistent with the formation of new dust reported for SN\,2018evt \citep{Wang2024}, though the lack of spectral information prevents the drawing of firm conclusions.

\subsection{Limits from spectroscopic and radio follow-up}
\label{subsec:multiwavelength}
\begin{figure*}
    \centering
    \includegraphics[width=\linewidth]{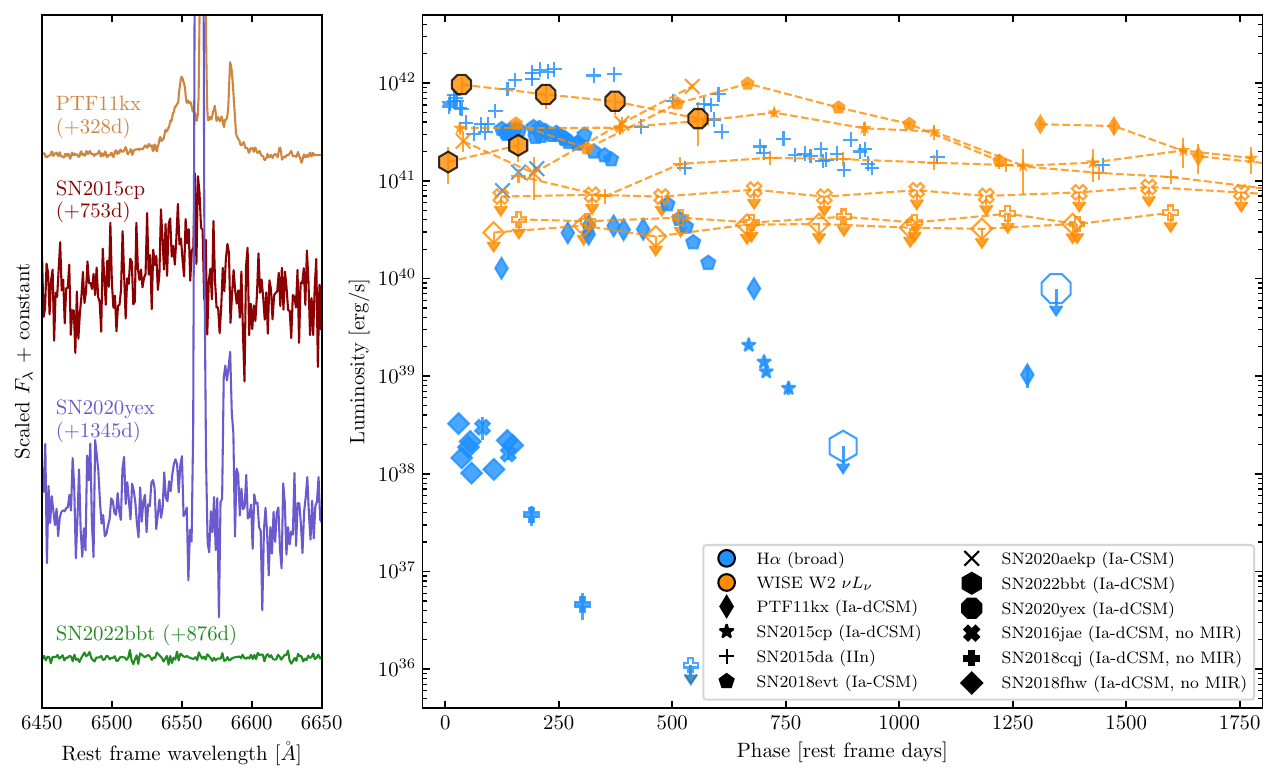}
    \caption{Optical spectroscopic and mid-infrared photometric signatures of CSM interaction.
    \textit{Left:} Spectra of the known delayed CSM-interacting SNe PTF11kx \citep{2017ApJ...843..102G} and SN\,2015cp \citep{2019ApJ...871...62G}, compared to spectra we obtained for SN\,2020yex and SN\,2022bbt, zoomed into the region around H$\alpha$. Both  H$\alpha$ lines ($>1000$ km\,s$^{-1}$) from CSM interaction, as well as unresolved host galaxy lines, are visible for PTF11kx and SN\,2015cp.
    In contrast, only narrow lines from the host galaxy are visible for SN\,2020yex, while no trace is detected in SN\,2022bbt.
    \textit{Right:} Comparison of the evolution of mid-infrared (\textit{NEOWISE}; orange) and H$\alpha$ (blue) luminosity for SNe\,Ia displaying delayed CSM interaction (labeled Ia-dCSM in the legend). We also show the corresponding luminosity evolution for SN\,2015da (SN IIn; \citealt{Tartaglia2020}), SN\,2018evt (SN\,Ia-CSM; \citealt{Wang2024}), SN\,2020aekp (SN\,Ia-CSM; \citealt{Sharma2023}) and three SNe\,Ia which show late-time nebular H$\alpha$ emission (SN\,2016jae; \citealt{2021A&A...652A.115E}, SN\,2018cqj; \citealt{2020ApJ...889..100P}, and SN\,2018fhw; \citealt{2019MNRAS.486.3041K, 2019MNRAS.487.2372V}). We show 5$\sigma$ upper limits on the broad/intermediate H$\alpha$ line (see Appendix~\ref{app:obs}) for two events in our sample in blue as the empty hexagon and square respectively.}
    \label{fig:halpha_mir_phase}
\end{figure*}
We obtained follow-up spectroscopy of SN\,2020yex and SN\,2022bbt (described in Appendix~\ref{app:obs}) with the goal of identifying broad or intermediate H or He lines typically seen in CSM-interacting SNe \citep{Smith2017_handbook, 2017ApJ...843..102G, 2019ApJ...871...62G}. The left panel of Figure~\ref{fig:halpha_mir_phase} show the spectra extracted at the expected position of the SN. While we detect narrow emission lines from the host galaxy for SN\,2020yex, we do not detect any broad or intermediate emission lines from either source, as seen in late-time spectra of PTF11kx and SN\,2015cp. We derive 5$\sigma$ upper limits on the H$\alpha$ luminosity of $L_{{\rm H}\alpha, +1345 {\rm d}} < 7.5 \times 10^{39}$\,erg\,s$^{-1}$ for SN\,2020yex and $L_{{\rm H}\alpha, +877 {\rm d}} < 1.9 \times 10^{38}$\,erg\,s$^{-1}$ for SN\,2022bbt (see Appendix \ref{app:obs}). To interpret these limits, we compare them to the evolution of the H$\alpha$ luminosity for PTF11kx and SN\,2015cp, as well as with a SN\,IIn, known SNe\,Ia-CSM and other low-luminosity SNe\,Ia which show delayed CSM interaction via H$\alpha$ emission.

The right panel of Figure~\ref{fig:halpha_mir_phase} shows that while the limits on SN\,2022bbt are consistent with the fast H$\alpha$ decline of SN\,2015cp \citep{2019ApJ...871...62G}, the non-detection rules out long-lived H$\alpha$ emission such as that seen in PTF11kx \citep{2013ApJ...772..125S, 2017ApJ...843..102G} and in SNe\,IIn \citep{Tartaglia2020}. Our non-detection of SN\,2020yex rules out SN\,IIn-like long-lived H$\alpha$ emission. We additionally show the luminous MIR emission for these SNe in Figure~\ref{fig:halpha_mir_phase}. In contrast to H$\alpha$ emission which fades on $1-2$\,yr timescales for most interacting SNe\,Ia, the MIR emission is much longer-lived, lasting thousands of days. Compared to SNe\,Ia-CSM and SNe\,Ia showing delayed CSM interaction in the MIR, the population of events of subluminous SNe\,Ia with late-time H$\alpha$ emission (SN\,2016jae, SN\,2018cqj, and SN\,2017fhw) exhibit H$\alpha$ luminosities which are $\sim 1000\times$ fainter at similar phases. If the X-ray or UV radiation powering the H$\alpha$ emission is similarly weak in these objects, the corresponding MIR emission would not be detectable, consistent with our non-detections in \textit{NEOWISE} data (Figure~\ref{fig:halpha_mir_phase}).

In our VLA follow-up observation of SN\,2022bbt at 932 days post-explosion, we did not detect a radio source, with a 5$\sigma$ flux density limit of $F_{\nu}<31\,\mu$Jy at a frequency of 5.5 GHz. Adopting the formalism of \cite{Chomiuk_2016} for $p = 3$, we infer the density of the environment assuming both ISM ($\rho_{csm}\propto$ constant) and wind-like ($\rho_{csm}\propto r^{-2}$) environments. We assume we are in equipartition with shock microphysical parameters with $\epsilon_e = \epsilon_B = 0.1$. For SN\,2022bbt, we constrain the mass loss rate for a wind-like environment to be $\dot{M} < 2\times10^{-6}$\,M$_\odot$\,yr$^{-1}$ for a wind velocity of $v_w = 10$\,km\,s$^{-1}$. We constrain the density for an ISM environment to be $n < 200$\,cm$^{-3}$. 

We can similarly use our limits on the intermediate-width H$\alpha$ luminosity $L_{{\rm H}\alpha}^{\rm Inter}$ to constrain the mass loss rate, following Equation 2 in \citet{Salamanca1998} which states
\begin{equation}
    L_{{\rm H}\alpha}^{\rm Inter} = \frac{1}{4} \epsilon_{{\rm H}\alpha} \frac{v_s^3}{v_w} \dot{M},
\end{equation}
where $v_s$ is the shell ejecta velocity, $v_w$ is the wind velocity, $\dot{M}$ is the mass loss rate, and $\epsilon_{{\rm H}\alpha}$ is an efficiency factor. Using $v_s = $ 10\,000\,km\,s$^{-1}$ for the shell ejecta, $v_w = $ 10\,km\,s$^{-1}$ for the wind, and a fiducial value of $\epsilon_{{\rm H}\alpha} = 0.01$, we find a mass loss limit of $\dot{M} < 4.8 \times 10^{-5}$\,M$_\odot$\,yr$^{-1}$ for SN\,2020yex and $\dot{M} < 1.2 \times 10^{-6}$\,M$_\odot$\,yr$^{-1}$ for SN\,2022bbt. We note that the efficiency factor $\epsilon_{{\rm H}\alpha}$ is a poorly constrained value that is estimated to be $\approx 0.1$ at early times, but eventually drops to zero \citep{Salamanca1998}. We choose $\epsilon_{{\rm H}\alpha} = 0.01$ since our measurements are at very late times. These limits are consistent with the radio measurements presented above.

\section{Discussion and summary}
\label{sec:discussion}

Our sample of five SNe with MIR signatures of delayed CSM interaction nearly doubles the previously known population, and in particular, substantially increases the diversity of SN\,Ia sub-types known to exhibit late CSM interaction. We can roughly estimate the fraction of SNe\,Ia showing late-time MIR brightening by considering SN\,2021jun, which was included in the flux-limited ZTF Bright Transient Survey (BTS; spectroscopically complete to 18.5~mag; \citealt{Fremling2020, Perley2020}). We restrict the BTS sample to SNe classified between 2018 June to 2021 June that pass BTS sample cuts \citep{Perley2020}, to allow for at least 1.5 years of \textit{NEOWISE} observations after the discovery of the SN. This results in a sample of 1466 SNe\,Ia, out of which we find one exhibiting late-time MIR brightening. We therefore set a lower limit of 0.05\% on the fraction of SNe\,Ia which exhibit delayed CSM interaction within 1.5\,yr of explosion.\footnote{This is a lower limit due to our restrictive filtering criteria, which may have excluded interacting SNe\,Ia that are coincident with complex galaxy backgrounds, active galactic nuclei hosts, or where the brightening was missed in the \textit{NEOWISE} observing gaps. Additionally, SN\,2021jun is a member of the underluminous Iax sub-class, indicating a potentially larger rate among SNe\,Iax compared to the broader SNe\,Ia population.} This is comparable with the rate of SNe\,Ia-CSM found by \citet{Silverman2013}, \citet{Dilday2012}, and \citet{Sharma2023} via searches around peak-light, indicating a similar fraction of SNe\,Ia exhibiting delayed interaction with CSM shells at $\sim 10^{16} - 10^{17}$\,cm. This is also consistent with the upper limits on rates of late-onset CSM interaction estimated from GALEX data by \citet{Dubay2022}.

The presence of pre-existing dust disfavors the classical double-degenerate progenitor model for SNe\,Ia, which are usually assumed to occur in clean environments \citep{Maoz2014, Taubenberger2017}. Instead, the existence of (potentially multiple) detached CSM shells indicate recent intense mass loss episodes that ended shortly prior to the explosion, which could arise in either a single degenerate scenario (via recurrent novae; \citealt{1999ApJ...522..487H, 2006Natur.442..276S, Dilday2012, 2012ApJ...761..182M, 2014MNRAS.443.1370D}), or a core-degenerate model with variable mass-loss rates during common-envelope (CE) ejection \citep{Livio2003, Soker2013, 2023MNRAS.521.4561S}. The dust mass lower limits inferred from photometry ($\sim 10^{-6} - 10^{-4}$\,M$_\odot$) suggest lower limits on total CSM masses of $\sim 10^{-4} - 10^{-2}$\,M$_\odot$ assuming a gas-to-dust mass ratio of $\approx 100$. On the high end, these lower limits on the masses may be similar to that estimated for PTF11kx \citep{Dilday2012}, which was argued by \citet{Soker2013} to be too massive to be explained by recent nova shells, though nova shells may be suitable for situations with lower CSM masses. Similarly, the delayed dynamical instability model by \citet{Han2006}, which invokes dynamical mass transfer at the time of explosion, cannot explain the time-delay between the explosion and the onset of CSM interaction.

Since our sample of SNe do not show signatures of interaction around the time of spectroscopic classification at peak light, the closest CSM shells must be farther than the ejecta radius at that time. For a typical classification phase of $\sim 20$\,d after explosion and an ejecta velocity of 10\,000\,km\,s$^{-1}$, this sets a lower limit on the minimum CSM radius of $\gtrsim 10^{15}$\,cm. Since the interaction appears to have begun by $\sim 100 - 300$\,d after explosion (Figure~\ref{fig:dust}), we constrain the CSM to be located at $\sim 10^{15} - 2 \times 10^{16}$\,cm. If the gas-to-dust mass ratio is significantly lower than the $\approx 100$ assumed above, the total CSM mass may be explainable by recurrent novae. Given the minimum CSM radius of $\sim 10^{15}$\,cm and a nova shell velocity of $\sim 1000$\,km\,s$^{-1}$ \citep{Aydi2020}, the ejection of the last nova shell must have occurred $\sim 0.5 - 3$\,yr prior to explosion.

In the core-degenerate case that may better explain larger CSM masses, slower CE ejection velocities of 10\,km\,s$^{-1}$ \citep{Soker2019} indicate that the CE was ejected $\sim 50-300$\,yr prior to explosion, consistent with the constraint by \citet{Soker2019} from SN\,2015cp \citep{2019ApJ...871...62G}. Our derived rate for late CSM interaction, which we find to be similar to that of SNe\,Ia showing prompt CSM interaction (SNe\,Ia-CSM), support the claim in \citet{Soker2022CEEDTD} that the time delay distribution between CE ejection and explosion may be more likely to be constant than the $\propto t^{-1}$ rate that is predicted for the delay time since star formation \citep{Friedmann2018, Heringer2019}. \citet{Soker2019} predicts that $\approx 10\%$ of SNe\,Ia should experience ejecta–CSM interaction within 30\,yr. Applying a flat time delay distribution to this prediction suggests that $\approx 0.5\%$ of SNe\,Ia show CSM interaction within 1.5\,yr, similar to our lower limit of $\gtrsim 0.1\%$. In another single-degenerate model, \citet{Justham2011} suggests that a rapidly-rotating super-Chandrasekhar WD can remain stable until after its donor's H-rich envelope is exhausted and contracts or is ejected, leading to a delayed explosion with little-to-no observable interaction at early times, in line with our observations.

Over half of our sample is comprised of overluminous 91T-like and super-Chandra SNe\,Ia\footnote{PTF11kx and SN\,2015cp were also initially classified as SNe\,Ia-91T at early phases, as are most known SNe\,Ia-CSM at very early phases \citep{Phillips2024}.}, despite their intrinsic rarity---with 91T-like events and super-Chandra events representing only $\approx 12$\% and $0.8$\% of all SNe\,Ia respectively \citep{Dimitriadis2024}. Since the overluminous and long-lived nature of their optical light curves suggests that these SNe have progenitors that are more massive than those of typical SNe\,Ia, their connection to delayed CSM interaction presented here points at a potential double degenerate scenario where the merger and subsequent explosion happens very shortly after the end of the CE phase \citep{Scalzo2010, Raskin2014, Siebert2024}. The lower fraction of normal SNe\,Ia exhibiting delayed CSM interaction then may point to a longer delay from the end of the common envelope to the time of explosion, if they also arise from the double degenerate channel. 

The connection between a SN\,Iax and a potential helium star companion (in SN\,2012Z; \citealt{McCully2022}) suggests that the late interaction seen in the Type Iax SN\,2021jun could be with circumstellar He, in a scenario similar to the helium-rich CSM surrounding SN\,2020eyj \citep{Kool2023}. While \citet{Kool2023} favored a single-degenerate He star channel for SN\,2020eyj, \citet{2023MNRAS.521.4561S} suggested that it could also be explained by a core-degenerate scenario where the last common envelope was produced by expansion and subsequent ejection of the He-rich envelope. Future MIR observations with {\it JWST} that are sensitive to the total dust mass via MIR continuum measurements may be able to distinguish between these scenarios based on the differing predictions for the total CSM mass in these two cases \citep{Moriya2019}. In addition, our existing observations cannot rule out MIR emission purely from massive dust echoes with particular geometries; future {\it JWST} observations can also explore this possibility. While our search was aided by the (short) overlap between systematic optical SN searches and the {\it NEOWISE} survey, we demonstrate that the luminous and long-lived IR signatures of CSM interaction provide a powerful probe of the environments of SNe\,Ia---suggesting that ground-based IR surveys like {\it WINTER} \citep{Lourie2020, frostig2024winter} and {\it PRIME} \citep{Yama2023}, as well as space-based surveys like {\it NEO-Surveyor} \citep{Mainzer2023} and the {\it Roman} space telescope \citep{Rose2021} are poised to systematically reveal this population.

\section*{}
We would like to acknowledge useful discussions with Melissa Graham, Wenbin Lu, Ken Shen, Joel Johansson and Kiyoshi Masui.
We thank Kristin Chiboucas for her excellent support with Gemini observations.
G.~M. acknowledges the support of the National Science Foundation and the LIGO Laboratory. K. D. was supported by NASA through the NASA Hubble Fellowship grant \#HST-HF2-51477.001 awarded by the Space Telescope Science Institute, which is operated by the Association of Universities for Research in Astronomy, Inc., for NASA, under contract NAS5-26555. K.~D. acknowledges support from a MIT Kavli Institute Fellowship. D.~F.'s contribution to this material is based upon work supported by the National Science Foundation under Award No.~AST-2401779.

This work is based on observations obtained at the international Gemini Observatory, a program of NSF NOIRLab, which is managed by the Association of Universities for Research in Astronomy (AURA) under a cooperative agreement with the U.S. National Science Foundation on behalf of the Gemini Observatory partnership: the U.S. National Science Foundation (United States), National Research Council (Canada), Agencia Nacional de Investigaci\'{o}n y Desarrollo (Chile), Ministerio de Ciencia, Tecnolog\'{i}a e Innovaci\'{o}n (Argentina), Minist\'{e}rio da Ci\^{e}ncia, Tecnologia, Inova\c{c}\~{o}es e Comunica\c{c}\~{o}es (Brazil), and Korea Astronomy and Space Science Institute (Republic of Korea).
This work was enabled by observations made from the Gemini North telescope, located within the Maunakea Science Reserve and adjacent to the summit of Maunakea. We are grateful for the privilege of observing the Universe from a place that is unique in both its astronomical quality and its cultural significance.

This paper includes data gathered with the 6.5 meter Magellan Telescopes located at Las Campanas Observatory, Chile. 

The National Radio Astronomy Observatory is a facility of the National Science Foundation operated under cooperative agreement by Associated Universities, Inc.

\facilities{NEOWISE, Gemini:Gillett (GMOS), VLA,  Magellan:Baade (FIRE, FourStar), WINTER, PO:1.2m, ZTF, ATLAS, Gaia, Spitzer}

\software{\texttt{astropy} \citep{astropy:2013},
         \texttt{matplotlib} \citep{Hunter:2007},
         \texttt{numpy} \citep{harris2020array},
         \texttt{pandas} \citep{reback2020pandas, mckinney2010data},
         \texttt{pypeit} \citep{Prochaska2020},
         \texttt{scipy} \citep{2020SciPy-NMeth},
        }

\bibliography{typeIa_delayedCSM}{}
\bibliographystyle{aasjournal}

\appendix
\restartappendixnumbering

\section{Archival objects and new sample details}
\label{app:sample}

\begin{figure*}
    \centering
    \includegraphics[width=\linewidth]{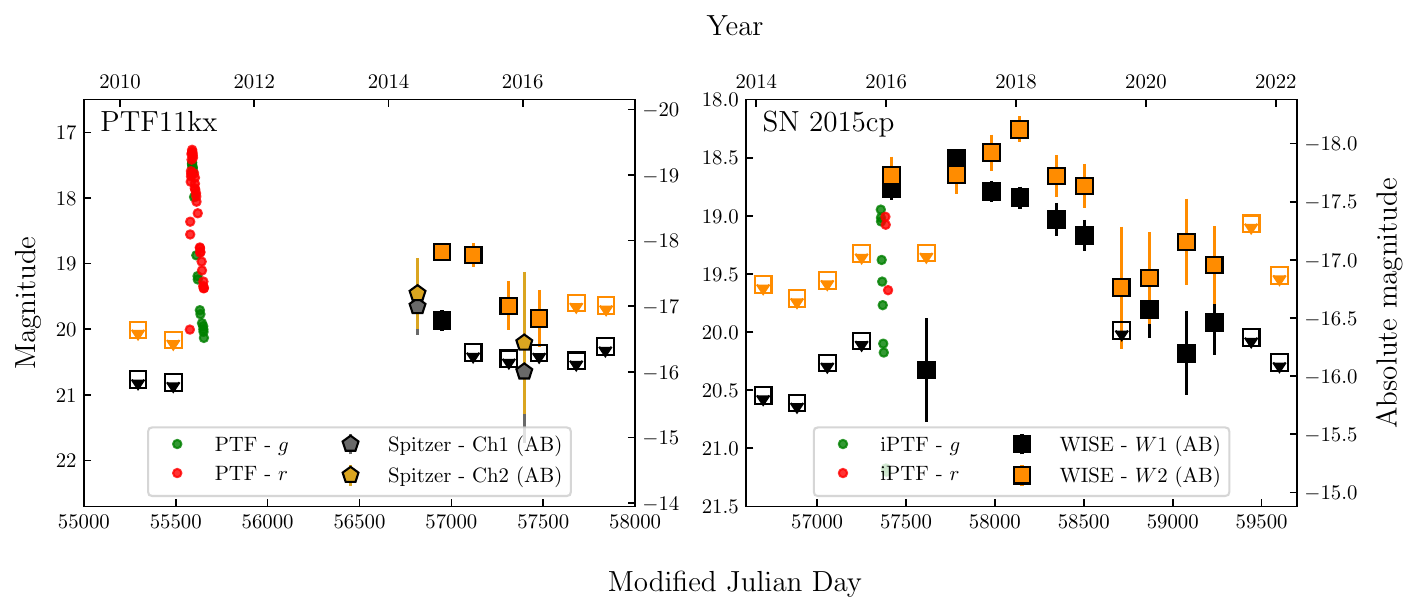}
    \caption{Optical and mid-infrared light curves of PTF11kx and SN\,2015cp. We present \textit{NEOWISE} photometry extracted using our difference imaging pipeline. We also show optical data from the Palomar Transient Factory (PTF; \citealt{Dilday2012}) and MIR data from \textit{Spitzer} for PTF11kx \citep{2017ApJ...843..102G}. The \textit{NEOWISE} and \textit{Spitzer} photometry are marginally offset, which can be attributed to differences in filters between each instrument \citep{2011ApJ...735..112J}. For SN\,2015cp, we show optical data from the intermediate PTF (iPTF; \citealt{2019ApJ...871...62G}).
    }
    \label{fig:old_lcs}
\end{figure*}

\begin{figure*}
    \centering
    \includegraphics[width=0.6\linewidth]{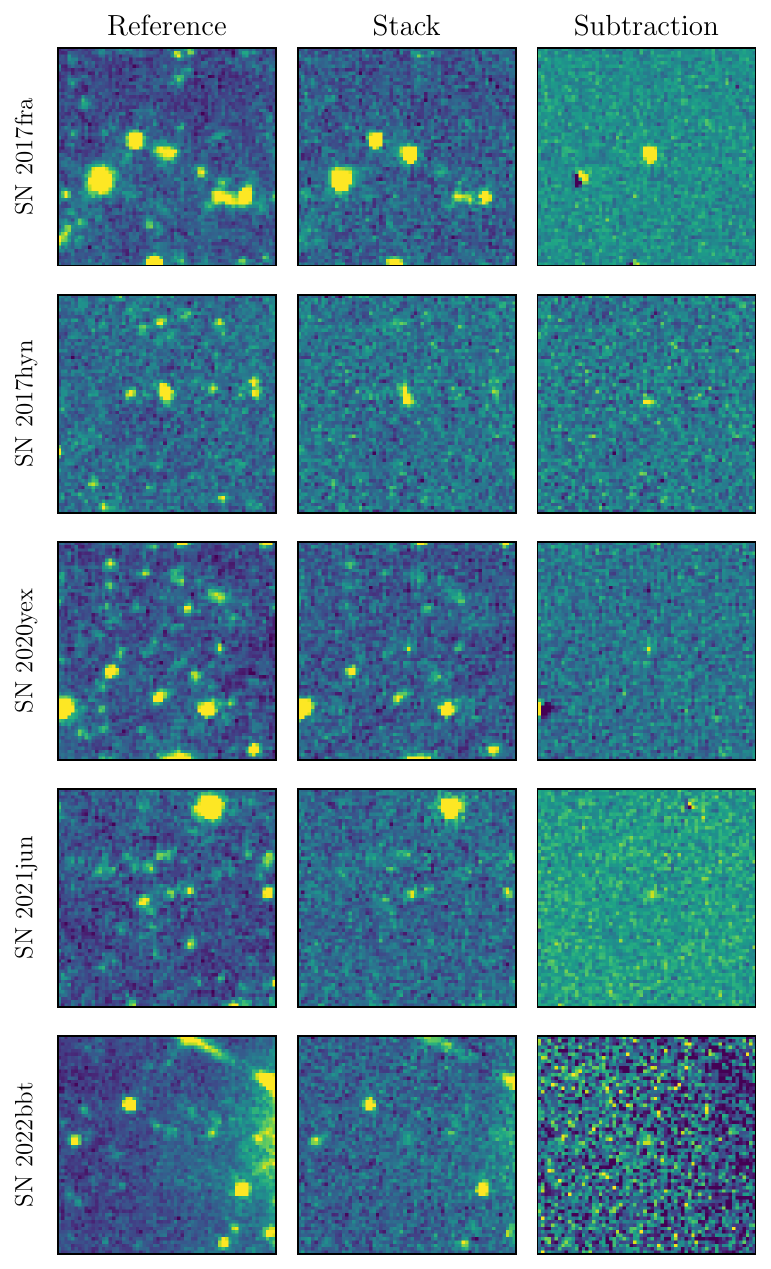}
    \caption{\textit{NEOWISE} W2 reference, stacked science, and subtraction cutouts of our sources at their brightest epoch. Each cutout has a width and height of 3$^\prime$ centered at the location of the transient, aligned north up and east left.
    }
    \label{fig:ref_stack_sub}
\end{figure*}

We show optical and MIR light curves of PTF11kx and SN\,2015cp in Figure~\ref{fig:old_lcs}, and details of our sample filtering in Table~\ref{tab:cand_cuts}. We also show \textit{NEOWISE} W2 reference, stack, and subtraction images in Figure~\ref{fig:ref_stack_sub}.

\begin{table}
    \centering
      \caption{Number of candidates for MIR-identified delayed CSM interaction removed at each stage of the filtering process.}
        \begin{tabular}{lll}
    \toprule
      Filtering criterion & Filtered candidates & Remaining candidates \\
          \midrule
          \midrule
    Initial sample from the Transient Name Server &   & 8572 \\
    No detections prior to the optical SN onset and no obvious subtraction artifacts & 8434 & 138  \\
    At least two \textit{NEOWISE} epochs ($\approx$ 1\,yr) and rising MIR emission & 120 & 18  \\
    Not a known SN\,Ia-CSM or SN\,Ia with delayed CSM interaction & 12 & 6\\
    \textit{NEOWISE} imaging not from temporally coincident nuclear activity & 1 & 5\\
          \bottomrule
    \end{tabular}

    \label{tab:cand_cuts}
\end{table}

\paragraph{SN\,2017fra} 
SN\,2017fra/ASASSN-17jq was discovered by ASAS-SN and was also observed by ATLAS and Gaia \citep{2017TNSTR.802....1P}. \citet{2017ATel10601....1C} spectroscopically classified it as a SN\,Ia-91T at $-8$\,d. It resides in a late-type galaxy that may be interacting with a companion and is at a redshift of $z=0.03$, resulting in a projected offset of 1.3\,kpc. \citet{2021ApJS..252...32J} first identified a connection between a MIR excess in its host and SN\,2017fra, reporting single-epoch dust fits. This MIR emission was also identified by \citet{Thevenot21RNAAS}, who also noticed that SN\,2017fra has a sibling in SN\,2016cda \citep{2024MNRAS.527.8015K}.\footnote{SN\,2016cda was the original candidate which appeared in our search, but the temporal coincidence of the MIR emission with SN\,2017fra makes it the more likely association.} In addition to the MIR data presented in \citet{2021ApJS..252...32J}, we show five additional \textit{NEOWISE} detection epochs in Figures \ref{fig:mir_lcs} and \ref{fig:lcs}. The MIR emission rises and fades smoothly over $\approx 5$\,yrs, peaking at $M_{\rm Vega, W2} \approx -22.3$ within $\approx 1.5$\,yrs after explosion; SN\,2017fra is the most luminous MIR SN\,Ia (Figures~\ref{fig:mir_lcs} and \ref{fig:lcs}).

\paragraph{SN\,2017hyn} 
SN\,2017hyn was discovered by ATLAS and also observed by Gaia \citep{2017TNSTR1236....1T}.
It was spectroscopically classified as a Ia-91T by \citet{2017TNSCR1251....1O} and is hosted in an edge-on disk galaxy at $z=0.053$, for a projected offset of 8.3\,kpc. \citet{Thevenot21RNAAS} first noted a \textit{NEOWISE} detection coincident with the optical peak, which is observed shortly after peak light. The subsequent long-lived MIR emission is similar to that of SN\,2017fra, exhibiting a sharp peak at 1.5 years after explosion.

\paragraph{SN\,2020yex}
SN\,2020yex was discovered by Gaia \citep{2020TNSTR3256....1H} and classified as a Ia-91T by \citet{2020TNSAN.211....1G}. It resides in a blue spheroidal galaxy with no distinct structure at $z=0.087$ and is at a projected offset of 2.7\,kpc. The MIR emission was first identified by \citet{Thevenot21Astronote}, who noted that the $W1$ luminosity was consistent with that of SNe\,Ia-CSM. Unlike the other MIR light curves in our sample, the \textit{NEOWISE} $W2$ light curve is consistent with a long-lived monotonic decline but the $W1$ light curve exhibits a peak at $\approx 1.5$\,yr after the explosion.

\paragraph{SN\,2021jun}
SN\,2021jun was discovered by ZTF \citep{2021TNSTR1208....1M} as part of its Bright Transient Survey, also observed by ATLAS and Pan-STARRS, and classified by \citet{2021TNSAN.159....1S, 2021TNSCR1619....1S} as a SN\,Iax. Its host is blue and spheroidal at $z=0.04$, with the SN at a projected offset of $<1$\,kpc. \textit{NEOWISE} observations reveal brightening MIR emission six months after the SN, coincident with a faint rebrightening in the ZTF photometry. The MIR light curve color changes dramatically $\approx 1$\,yr later, becoming significantly more blue, after which it reddens again.

\paragraph{SN\,2022bbt} 
SN\,2022bbt was discovered by ATLAS \citep{2022TNSTR.255....1T} and also observed by Pan-STARRS, Gaia, and ZTF in its Bright Transient Survey. It is offset ($7.3^{\prime\prime}$, 7.9\,kpc projected distance at $z = 0.049$) from a face-on spiral host. \citet{2022TNSCR.344....1T} classify it as a SN\,Ia, while \citet{2022TNSCR1640....1J} classify it as a Ia-pec. Using the spectrum from \citet{2022TNSCR.344....1T} and the supernova fitting code NGSF \citep{2022TNSAN.191....1G}, we find evidence for low velocity C\,II lines, which together with the luminous (peaking at $M\approx -20$\,AB mag in $r$-band) light curve, is consistent with SN\,2022bbt being a super-Chandrasekhar (09dc-like) SN\,Ia \citep{Howell2006}. Since this is most recent event in our sample, only one \textit{NEOWISE} detection is available; however, the luminosity is consistent with the rest of the population (Figure~\ref{fig:mir_lcs}).

\section{Additional follow-up observations}
\label{app:obs}

We acquired Gemini GMOS-N \citep{2004PASP..116..425H} spectra on UT 2024-08-07 (SN\,2022bbt) and Gemini GMOS-S spectra on UT 2024-10-26 (SN\,2020yex) using the B480 gratings and $2 \times 2$ binning on the Hamamatsu detectors, resulting in 1.3\,\AA~spectral resolution, sufficient for distinguishing narrow host lines from the broad or intermediate lines indicative of SN ejecta. The sources were acquired using blind-offsets from nearby stars and each spectrum was integrated for 80\,min. The spectra were reduced using standard routines in \texttt{PypeIt} \citep{Prochaska2020}, including flat-fielding and wavelength calibration using arcs. Flux calibration was performed using a standard star observation followed by one-dimensional spectral extraction at the expected position of each SN. No broad or intermediate emission lines were identified in either spectrum (see Section \ref{subsec:multiwavelength}). For each source, we derived an upper limit on the intermediate emission line flux by rebinning the spectrum to the expected $\sim 20$\,\AA~width (FWHM $\approx 1000$\,km\,s$^{-1}$) of the intermediate emission line \citep{2017ApJ...843..102G}, then taking the standard deviation of a $\sim 300$~\AA~featureless section of the spectrum near the expected H$\alpha$ line. 

We additionally attempted infrared spectroscopy of SN\,2020yex with the FIRE \citep{Simcoe2013} spectrograph on the Magellan/Baade telescope on UT 2024-07-04 in the low-resolution prism mode. The source was acquired using a blind-offset from a nearby star. No trace was detected at the source position with a total integration time of $\approx 20$\,min. We acquired NIR imaging in $J$ and $Ks$ bands of SN\,2017fra on UT 2024-07-09 using the Fourstar \citep{2013PASP..125..654P} camera on the Magellan/Baade telescope. The dithered images (amounting to a total exposure time of $\approx 1200$\,s and $\approx 900$\,s in $J$ and $Ks$ bands, respectively) were stacked, astrometrically and photometrically calibrated using a custom pipeline \citep{De2020}. No source was detected at the SN position in the stacked images down to a $5\sigma$ limiting magnitude of $> 20.5$\,Vega\,mag and $>19.0$\,Vega mag in $J$ and $Ks$ bands, respectively. We caution that these limits are derived from aperture photometry (together with visual inspection) at the SN position since there are no archival images of the quiescent host galaxy at this depth.

We observed SN\,2021jun in the NIR $J$-band with the {\it WINTER} camera \citep{Lourie2020, frostig2024winter} mounted on the Palomar 1-m telescope on UT 2024-05-12, with a total exposure time of 3600\,s. The images were processed using the {\it WINTER} data reduction pipeline \texttt{mirar} \citep{mirar_doi} and image subtraction was performed relative to $J$-band images from the UKIRT Hemisphere Survey \citep{Dye2018}. No sources were detected in difference imaging at the location of SN\,2021jun, with a 5$\sigma$ limiting magnitude of $J \sim 20.1$ (AB).

We also searched for radio emission from SN\,2022bbt using the VLA \citep{1998AJ....115.1693C} on UT 2024-08-28 in C-band, with the array in B-configuration ($1^{\prime\prime}$ beamwidth, to distinguish between host galaxy nuclear emission). We used 3C286 as a flux calibrator, while using J1516+1932 as a phase calibrator. The data was calibrated with the VLA CASA Calibration Pipeline using CASA 6.5.4. and imaged manually with the \texttt{TCLEAN} task in CASA, using natural weighting. No radio source was detected, with a 5$\sigma$ flux density limit of $F_{\nu}<31\,\mu$Jy.

\end{document}